\documentclass[superscriptaddress,twocolumn,showpacs,aps]{revtex4}

\usepackage{graphicx}% Include figure files
\usepackage{dcolumn}% Align table columns on decimal point
\usepackage{bm}% bold math

\renewcommand{\vec}[1]{\mbox{\boldmath $#1$}}

\begin{document}

\preprint{}

\title{Perturbative HFB model for many-body pairing correlations}

\author{K. Hagino}
\affiliation{
Department of Physics, Tohoku University, Sendai 980-8578, Japan}

\author{H. Sagawa}
\affiliation{
Center for Mathematical Sciences, University of Aizu, Ikki-machi, 
Aizu-Wakamatsu, Fukushima 965-8580, Japan}

\date{\today}% It is always \today, today,

\begin{abstract}
We develop a perturbative model to treat 
the off-diagonal 
components in the Hartree-Fock-Bogoliubov (HFB) transformation matrix,
which are neglected in the BCS approximation.  
Applying the perturbative model to a weakly bound 
nucleus $^{84}$Ni, it is shown that the  
perturbative approach reproduces well the solutions of the HFB method 
both for the quasi-particle energies and the radial dependence 
of quasi-particle wave functions. 
We find that the 
non-resonant part of the continuum single-particle state can 
acquire an appreciable occupation probability when there exists 
a weakly bound state close to the Fermi surface. 
This result originates from 
the strong coupling between the continuum 
particle state and the weakly bound state,
and is absent in the BCS approximation.  The limitation of 
the BCS  approximation is pointed out   
in comparison with the HFB and the present perturbative model.
\end{abstract}

\pacs{21.60.-n,21.60.Jz,21.10.Pc}

\maketitle

\section{Introduction}

The pairing is the most important correlation beyond the 
independent particle approximation in the nuclear mean field 
\cite{RS80,DHJ03}. 
For the description of nuclei close to the beta stability line, 
the BCS approximation has been successfully employed in order to 
take into account the pair correlation. 
Although this approximation 
does not treat the interplay between the particle-hole
and the pairing channels in a fully consistent manner \cite{RS80}, 
one can expect that the defect is minor in well bound nuclei. 
As the Fermi energy approaches to zero, however, 
virtual scattering of nucleon pairs to the positive energy spectra 
may significantly alter the situation. 
If the continuum states acquire an appreciable occupation probability, 
in the BCS approximation, 
the particle density does not tend to zero asymptotically. 
In this case, the unphysical particle gas surrounding 
the nucleus appears \cite{DFT84,DNW96,BDP99}, and one has to use 
the Hartree-Fock-Bogoliubov (HFB) method 
\cite{RS80,DFT84,DNW96,BDP99,B00,BHR03,GSGL01,HM03}, 
which is 
theoretically 
more robust than the BCS approximation. 

Recently, Sandulescu {\it et al.} have argued that 
the non-resonant part of the single particle 
continuum spectra can be neglected 
in the pairing correlation 
\cite{GSGL01,SGL00,SGTH03} (see also Refs. \cite{BEP96,KHL01}). 
Since the wave function for a resonance state is well localized inside
the nucleus, the unphysical particle gas is negligible 
if the non-resonant part is excluded from the 
positive energy space. 
By taking only a few narrow resonance states into account, the authors
of Refs. \cite{GSGL01,SGL00,SGTH03,KHL01} have demonstrated that the 
BCS approximation leads to similar results to the HFB method 
concerning the binding energy as well as the root mean square radius 
both for stable and weakly bound nuclei. 

Those results, however, do not necessarily mean that the BCS
approximation provides a good representation for quasi-particle 
wave functions. 
In fact, 
Dobaczewski {\it et al.} have shown \cite{DNW96} 
that the smaller  
part of the two component 
quasi-particle wave functions in the BCS approximation
behaves quite differently from that obtained with the HFB method,
although the larger component is similar to each other. 
Since transitions are usually more sensitive to the details of wave
function than energies, a more intriguing test of the BCS
approximation should be for 
quasi-particle wave functions. 

The purpose of this paper is to discuss 
how the BCS approximation can be improved while 
retaining its intuitive picture. To this end, 
we start from the full HFB equations and simplify them by 
the BCS approximation. We then propose a model which 
treats the effect beyond the BCS
approximation by the perturbation theory. 
We perform detailed comparisons between  the three models, 
the BCS,  the HFB and the perturbative approximations for 
the quasi-particle wave function in 
weakly bound nuclei, and clarify the limitation of 
the applicability of the BCS approximation.

The paper is organized as follows. In Sec. II, we give the HFB
equations in the coordinate space, and formulate the perturbative 
approach to HFB. In Sec. III, we apply the perturbative approach to 
neutron states in the weakly bound nucleus $^{84}$Ni, and discuss 
the applicability of the perturbation formulas. 
The summary of the paper is then given in Sec. IV. 

\section{Perturbative approach to HFB}

Let us start with the HFB equations in the coordinate space
representation \cite{DFT84,DNW96,B00}, 
\begin{equation}
\left(
\begin{array}{cc}
\hat{h}-\lambda&\Delta(r) \\
\Delta(r)&-\hat{h}+\lambda
\end{array}
\right)
\left(
\begin{array}{c}
U_k(r)\\
V_k(r)
\end{array}
\right)
=E_k
\left(
\begin{array}{c}
U_k(r)\\
V_k(r)
\end{array}
\right),
\label{HFB}
\end{equation}
where 
$\hat{h}$ and $\Delta(r)$ are the mean-field Hamiltonian and 
the pairing potential, respectively. Here, we have assumed that the 
nucleon-nucleon interaction is a zero range force so that 
the mean-field and the pairing potentials are local. 
$\lambda$ and $E_k$ are the Fermi energy and the quasi-particle
energy, respectively. 
In order to solve the HFB equations (\ref{HFB}), we expand the 
quasi-particle wave functions $U_k(r)$ and $V_k(r)$ on the 
Hartree-Fock (HF) basis 
$\varphi_i$,
\begin{eqnarray}
U_k(r)&=&\sum_iu^{(k)}_{i}\,\varphi_i(r), 
\label{wf1} \\
V_k(r)&=&\sum_iv^{(k)}_{i}\,\varphi_i(r),\label{wf2}
\end{eqnarray}
where the wave function $\varphi_i$ satisfies 
$\hat{h}\varphi_i=\epsilon_i\varphi_i$. 
This leads to an eigenvalue problem of the matrix ${\cal H}_{\rm HFB}$ 
which is given by 
\begin{equation}
({\cal H}_{\rm HFB})_{ij}=\left(
\begin{array}{cc}
(\epsilon_i-\lambda)\delta_{i,j}&\Delta_{ij} \\
\Delta_{ij}&(-\epsilon_i+\lambda)\delta_{i,j}
\end{array}
\right), 
\label{HFB2}
\end{equation}
where 
\begin{equation}
\Delta_{ij}=\int dr\,\varphi^*_i(r)\Delta(r)\varphi_j(r). 
\end{equation}
This is the so called two-basis method for the HFB
equations\cite{GBD94,THB95,THF96,TFH97}. 

The BCS approximation is achieved by neglecting the off-diagonal
components of $\Delta_{ij}$ in Eq. (\ref{HFB2}) \cite{BHR03}. 
The resultant equations, 
\begin{eqnarray}
(\epsilon_i-\lambda)u^{(k)}_i+\Delta_{ii}\,v_i^{(k)}&=&E_k^{(0)}\,u_i^{(k)}, \\
\Delta_{ii}\,u_i^{(k)}+(-\epsilon_i+\lambda)v^{(k)}_i&=&E_k^{(0)}\,v_i^{(k)},
\end{eqnarray}
can be solved easily. The solutions are the well known BCS solutions 
\cite{RS80}, 
\begin{equation}
\left(
\begin{array}{c}
u_i^{(k)} \\v_i^{(k)} \end{array} \right)
=
\left(
\begin{array}{c}
u_k^{\rm BCS} \\ v_k^{\rm BCS} 
\end{array} \right)\,\delta_{i,k}\,, 
\label{BCS}
\end{equation}
with
\begin{eqnarray}
u_k^{\rm BCS}&=&\sqrt{\frac{1}{2}\left(
1+\frac{\epsilon_k-\lambda}{E^{(0)}_k}\right)}, \\
v_k^{\rm BCS}&=&\sqrt{\frac{1}{2}\left(
1-\frac{\epsilon_k-\lambda}{E^{(0)}_k}\right)}, \\
E^{(0)}_k&=&\sqrt{(\epsilon_k-\lambda)^2+\Delta_{kk}^2}. 
\end{eqnarray}
From Eqs. (\ref{wf1}),
(\ref{wf2}), and (\ref{BCS}), it is apparent that the quasi-particle
wave functions in the BCS approximation are given by 
\begin{eqnarray}
U_k(r)&=&u_k^{\rm BCS}\,\varphi_k(r), \\
V_k(r)&=&v_k^{\rm BCS}\,\varphi_k(r). 
\end{eqnarray}
Thus, the two components of the quasi-particle wave function are
simply proportional to each other in the BCS approximation
\cite{DNW96,B00,BHR03}. 

We next consider the effect of the off-diagonal components of
$\Delta_{ij}$. To this end, we rewrite the HFB Hamiltonian (\ref{HFB2}) as  
\begin{equation}
{\cal H}_{\rm HFB}={\cal H}_{\rm BCS}+\Delta{\cal H},
\end{equation}
with
\begin{eqnarray}
({\cal H}_{\rm BCS})_{ij}&=&
\left(
\begin{array}{cc}
(\epsilon_i-\lambda)&\Delta_{ii} \\
\Delta_{ii}&(-\epsilon_i+\lambda)
\end{array}
\right)\delta_{i,j}, \\
(\Delta{\cal H})_{ij}&=&
\left(
\begin{array}{cc}
0&\Delta_{ij}-\Delta_{ii}\delta_{i,j} \\
\Delta_{ij}-\Delta_{ii}\delta_{i,j}&0
\end{array}
\right). 
\end{eqnarray}
The eigen functions of 
${\cal H}_{\rm BCS}$ form the unperturbed basis, which is given by 
\begin{eqnarray}
|\phi_i^{(0)}\rangle=(0\cdots u_i^{\rm BCS}\cdots 0,0
\cdots v_i^{\rm BCS}\cdots 0)^T. 
\end{eqnarray}
Treating $\Delta{\cal H}$ by the standard perturbation theory, 
we find that the correction to the quasi-particle wave function and
the quasi-particle energy is given by, 
\begin{eqnarray}
|\phi_i\rangle &=& {\cal N}_i(|\phi_i^{(0)}\rangle+|\delta\phi_i\rangle), \\
&=&
{\cal N}_i\left[|\phi_i^{(0)}\rangle+
\sum_{j\neq i}\left(
\frac{\langle\phi_j^{(0)}|\Delta {\cal H}|\phi_i^{(0)}\rangle}
{E_i^{(0)}-E_j^{(0)}}\right)\,|\phi_j^{(0)}\rangle\right] , 
\end{eqnarray}
and 
\begin{equation}
E_i=E_i^{(0)}+\Delta E_i = E_i^{(0)}+
\sum_{j\neq i}
\frac{|\langle\phi_j^{(0)}|\Delta {\cal H}|\phi_i^{(0)}\rangle|^2}
{E_i^{(0)}-E_j^{(0)}},
\end{equation}
respectively, to the leading order. 
Here ${\cal N}_i$ is introduced to normalize the wave 
function $|\phi_i\rangle$. 
Notice that the first order correction to the quasi-particle energy
vanishes. The matrix element of 
$\Delta {\cal H}$ is given by, 
\begin{equation}
\langle\phi_j^{(0)}|\Delta {\cal H}|\phi_i^{(0)}\rangle 
=(\Delta_{ij}-\Delta_{ii}\delta_{i,j})\,(u^{\rm BCS}_j
v^{\rm BCS}_i+v^{\rm BCS}_ju^{\rm BCS}_i)\,. 
\end{equation}
We call the perturbative model the p-HFB. 

\section{Applicability of the perturbation formulas}

We now apply the perturbative approach, the p-HFB, to a drip-line
nucleus in the $^{84}$Ni region. 
Following Ref. \cite{HM03}, we take a Woods-Saxon form
for the mean-field and the pairing potentials, that is, 
\begin{eqnarray}
V_{\rm WS}(r)
&=&-V_0\,f(r)+V_{ls}\,\frac{1}{r}\frac{df}{dr}\,\vec{l}\cdot\vec{s} \\
\Delta(r)&=&\Delta_0\,f(r), \\
f(r)&=&\frac{1}{1+\exp[(r-R_0)/a]}. 
\end{eqnarray}
We use the parameters, $V_0$=38.5 MeV, $V_{ls}$=14 MeV$\cdot$ fm, $R_0$=
5.63 fm, and $a$=0.66 fm in order to simulate the neutron potential around
the $^{84}$Ni region. 
The strength of the pairing potential is determined so that the
average pairing gap defined as \cite{HM03} 
\begin{equation}
\bar{\Delta}=\frac{\int^\infty_0r^2dr\,\Delta(r)f(r)}
{\int^\infty_0r^2dr\,f(r)},
\end{equation}
is equal to 1.0 MeV. The Fermi energy $\lambda$ is an input parameter 
in this simplified model, and we take $\lambda = -0.5$ MeV in the
calculations shown below. 
For simplicity, the positive energy solutions of the 
mean-field Hamiltonian 
are treated by discretizing them with a box of 30 fm. 

\begin{figure}
\includegraphics[clip,scale=0.5]{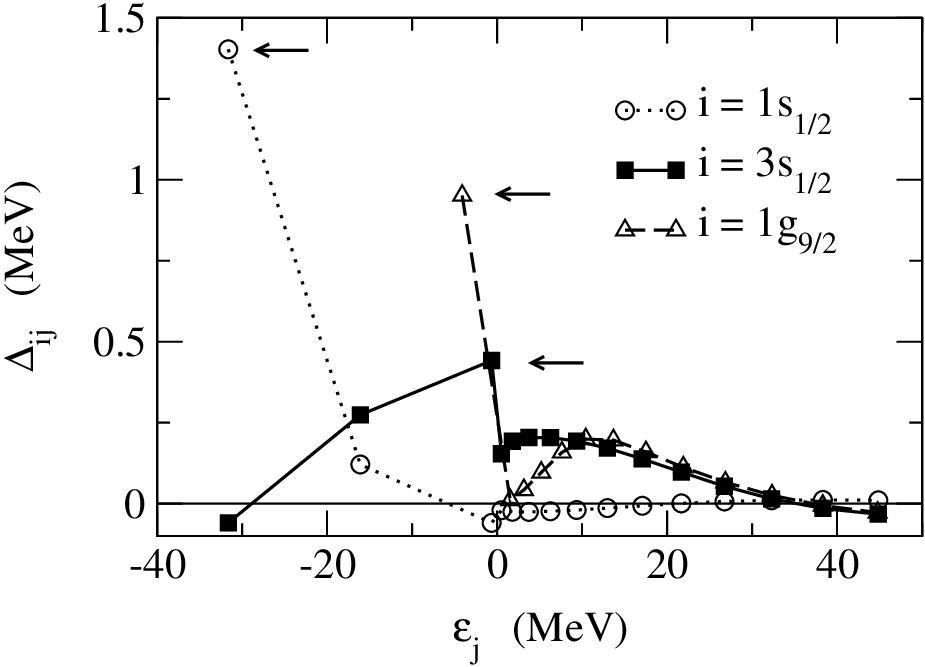}
\caption{
The matrix elements of the pairing potential $\Delta(r)$. 
These are plotted as a function of the single-particle energy 
$\epsilon_j$ for 
the state $|\varphi_j\rangle$ with a fixed bound state
$\langle\varphi_i|$, 
which has the same total and orbital
angular momenta $j$ and $l$ as the state $\varphi_j$. 
The open circles, the filled squares, and the open triangles are 
for the neutron 1$s_{1/2}$ state at $-$31.6 MeV, the 
3$s_{1/2}$ state at $-$0.634 MeV, and the 
1$g_{9/2}$ state at $-$4.11 MeV, respectively. The diagonal components 
are denoted by the arrows. 
}
\end{figure}

We first discuss 
the matrix elements $\Delta_{ij}$ of the pairing potential
$\Delta(r)$ on the Hartree-Fock basis.  
Figure 1 shows 
the matrix elements 
as a function of the single-particle energy of the state $j$ 
for a fixed state $i$. 
For well bound states, such as the 1$s_{1/2}$
state at $\epsilon=-31.6$ MeV, they are all close to
zero except for the diagonal component, which is denoted by the 
arrow (see the open circles). This justifies the success of the BCS
approximation for stable nuclei. For weakly bound states, in contrast, 
the off-diagonal components are the same order as the diagonal 
component. Specifically, the coupling between the weakly bound state 
and the positive energy states persist up to about 30 MeV. 
This is the case both for the $s_{1/2}$ state (the filled square) 
and for a state with larger 
angular momentum such as the $g_{9/2}$ state (the open triangle). 
Since the BCS approximation neglects completely the off-diagonal
components of the pairing potential, a careful investigation of the 
effects of the off-diagonal components is therefore 
necessary for weakly bound nuclei. 

\begin{table}[hbt]
\caption{The quasi-particle energy $E$ and the occupation probability
$v^2$ for the four lowest neutron s$_{1/2}$ quasi-particle 
states
obtained with several methods. For the BCS method, the corresponding 
single particle level and the single particle energy are also tabulated. }
\begin{center}
\begin{tabular}{c|cccc}
\hline
\hline
 & $E$ (MeV) & $v^2$ & s.p. state & $\epsilon$ (MeV) \\
\hline
HFB & 0.421 & 0.553 &             &                  \\
    & 1.045 & 0.039 &            &                  \\
    & 2.327 & 0.010 &            &                  \\
    & 4.259 & 0.003 &            &                  \\
\hline
BCS & 0.462 & 0.645 & 3s$_{1/2}$  &  $-$0.634 \\
    & 1.010 & 0.007 & 4s$_{1/2}$  &  $+$0.509 \\
    & 2.305 & 0.004 & 5s$_{1/2}$  &  $+$1.803 \\
    & 4.244 & 0.002 & 6s$_{1/2}$  &  $+$3.742 \\
\hline
p-HFB & 0.400 & 0.606 &   &   \\
             & 1.039 & 0.033 &   &   \\
             & 2.318 & 0.005 &   &   \\
             & 4.250 & 0.001 &   &   \\
\hline
\hline
\end{tabular}
\end{center}
\end{table}

Let us now solve the HFB equations by diagonalizing the HFB Hamiltonian
(\ref{HFB2}) and compare with the BCS approximation as well as with the
p-HFB method discussed in the previous section. 
We include the single-particle levels up to $\epsilon$= 50 MeV. We
have checked that the results do not change significantly even when
higher lying states are included. 
Table 1 compares the HFB method with the BCS approximation for 
the quasi-particle energy $E$ and the occupation probability $v^2$ for
the four lowest neutron s$_{1/2}$ quasi-particle states. 
The occupation probability is defined in terms of the quasi-particle 
wave function as 
\begin{equation}
v_k^2=\int^\infty_0dr\,|V_k(r)|^2. 
\end{equation}
Also shown in the table is the 
corresponding single-particle state and its energy for the BCS approximation. 
We find that there is a good correspondence between the HFB
and the BCS solutions, especially for the quasi-particle energy. 
One can thus assign unambiguously 
a particular single particle state to each HFB state as an
unperturbative state. 

\begin{figure}
\includegraphics[clip,scale=0.5]{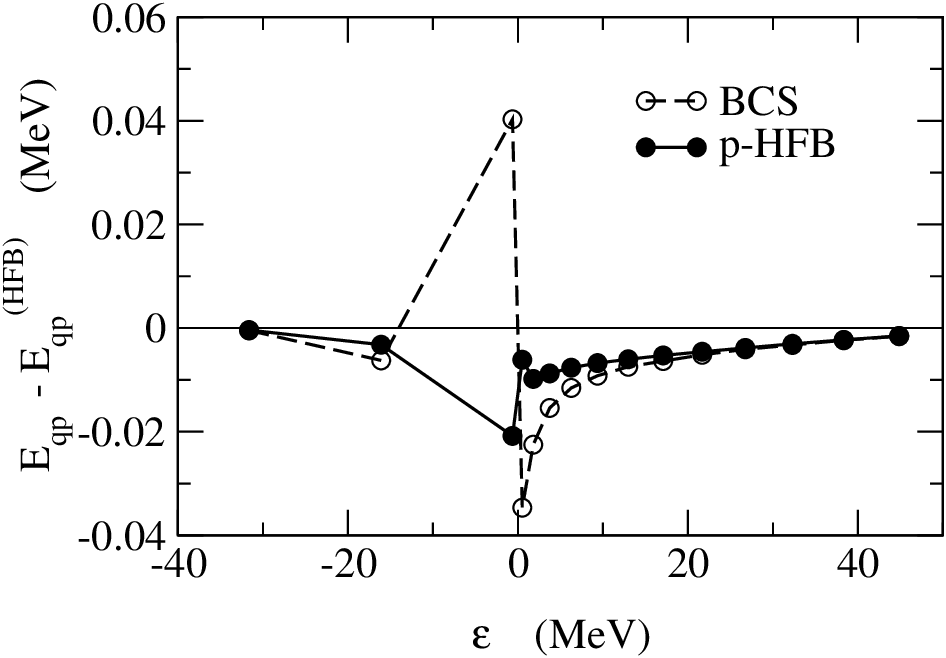}
\caption{
The deviation of the neutron s$_{1/2}$ quasi-particle energy, obtained 
with the BCS method (the dashed line) 
and the perturbative method p-HFB (the solid line),  
from the corresponding HFB energy as a 
function of the single-particle energy.}
\end{figure}

\begin{figure}
\includegraphics[scale=0.45,clip]{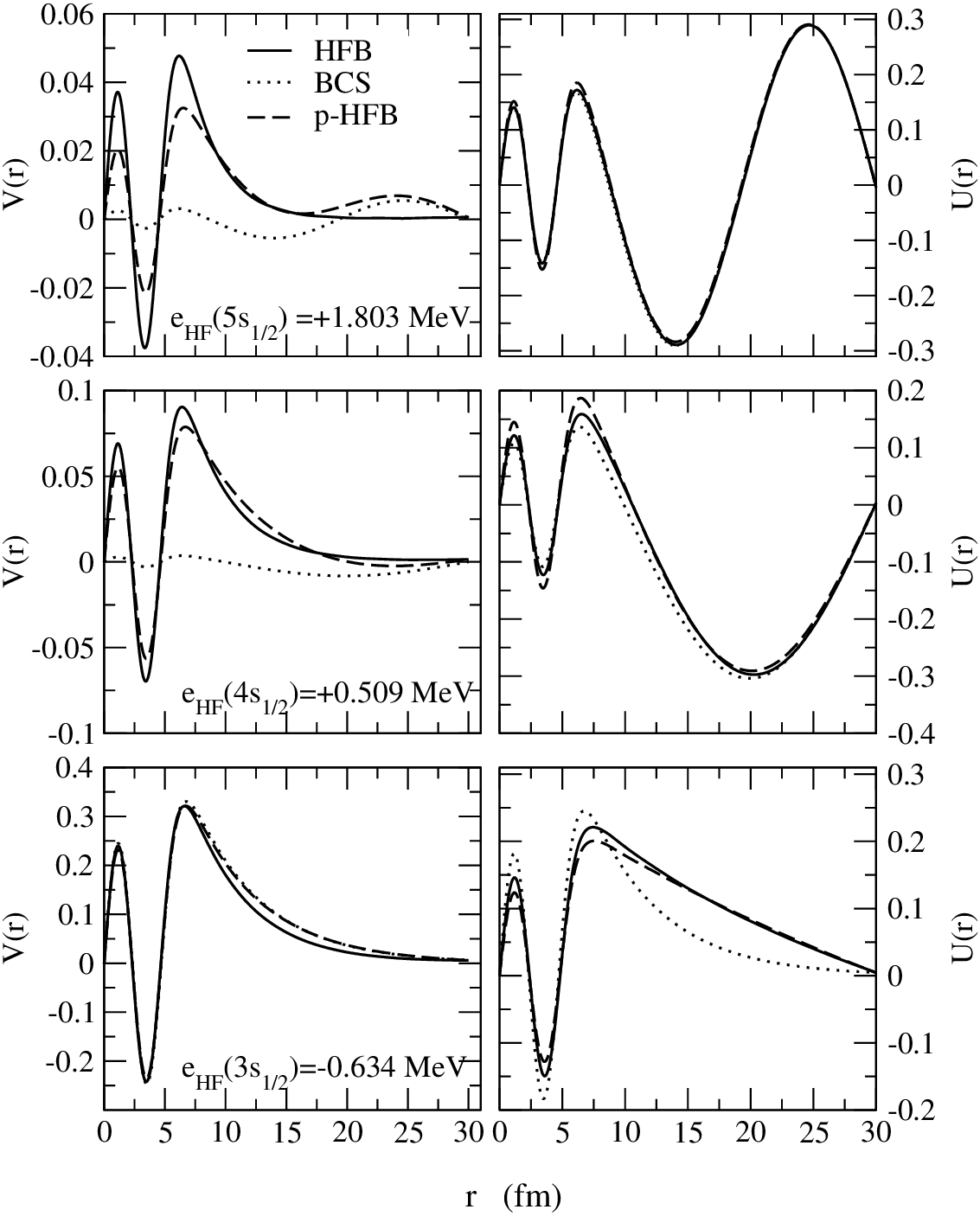}
\caption{
The quasi-particle wave functions for the neutron s$_{1/2}$ states 
obtained with 
the three methods, HFB, BCS, and p-HFB. 
The left and the right panels are for 
the lower ($V$) and the upper ($U$) 
components of the wave function, respectively. 
The solid line is the solution of the HFB method, while the dotted
line denotes the wave function in the BCS approximation. The result 
of the perturbative method p-HFB is shown by the dashed line. 
The top, middle, and bottom panels are for the 5s$_{1/2}$, 4s$_{1/2}$, 
and 3s$_{1/2}$ states, respectively. }
\end{figure}

Figure 2 shows the difference in the quasi-particle energy from the 
HFB energy for the neutron s$_{1/2}$ states as a
function of the energy of the corresponding
single-particle state. The dashed
and the solid lines are obtained with the BCS approximation and the 
perturbation theory, respectively. Surprisingly, 
the quasi-particle energy is off only by 0.04 MeV at most in 
the BCS approximation, and the 
approximation seems to work well as long as the quasi-particle
energy is concerned. This is in accordance with the finding in Refs. 
\cite{GSGL01,SGL00,SGTH03,KHL01}. The perturbative method (p-HFB)
further improves 
the result, and the difference in the quasi-particle energy becomes
even smaller (see the solid line). 
Figure 3 compares the quasi-particle wave functions for the first
three quasi-particle states. The left and the right panels show the
$V(r)$ and the $U(r)$ functions, respectively. 
The solid, the dotted, and the dashed lines denote the results of the
HFB method, the BCS approximation, and the p-HFB method, 
respectively. As noted by Dobaczewski {\it et al.} 
\cite{DNW96}, the smaller component of the wave function in the BCS 
approximation is inconsistent with the HFB wave function. 
Notice, however, that the perturbative method p-HFB dramatically improves 
the radial dependence of the wave function, suggesting that the BCS 
approximation still provides a good starting point to begin with. 
For the larger component, although the difference between the HFB and the BCS
wave functions is small (but see below for the wave function for the 
1g$_{9/2}$ state), the perturbation again improves the shape of the 
wave function in a consistent way. 
The results of the p-HFB method are summarized in Table 1. 

\begin{figure}
\includegraphics[scale=0.5,clip]{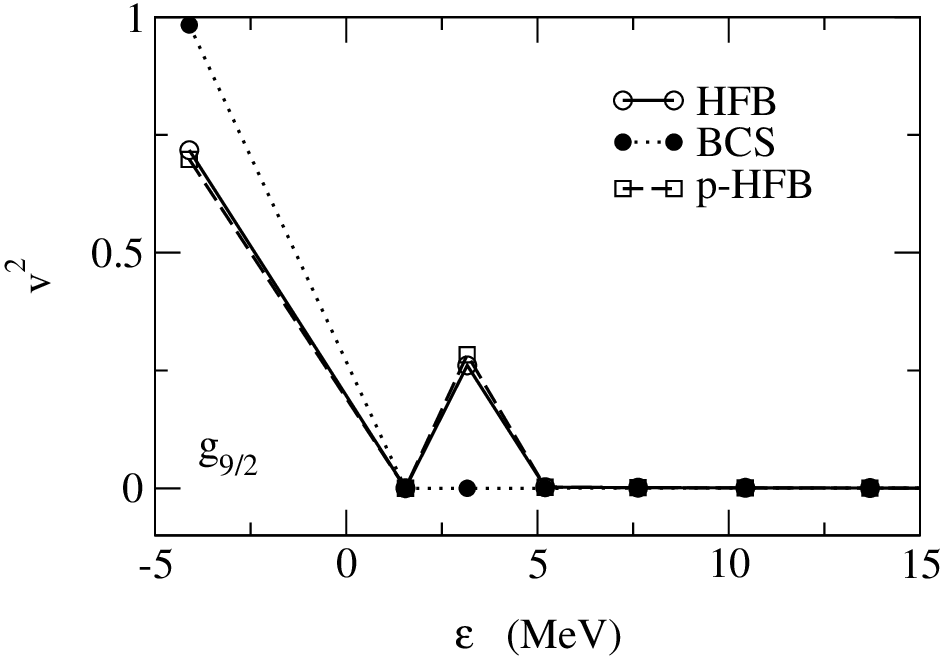}
\caption{The occupation probability for the $g_{9/2}$ states 
as a function of the corresponding HF energy $\epsilon$. The solid 
and the dotted lines are results of the HFB method and the BCS approximation, 
respectively, while the dashed line is obtained by the 
perturbative method, p-HFB.}
\end{figure}

\begin{figure}
\includegraphics[scale=0.45,clip]{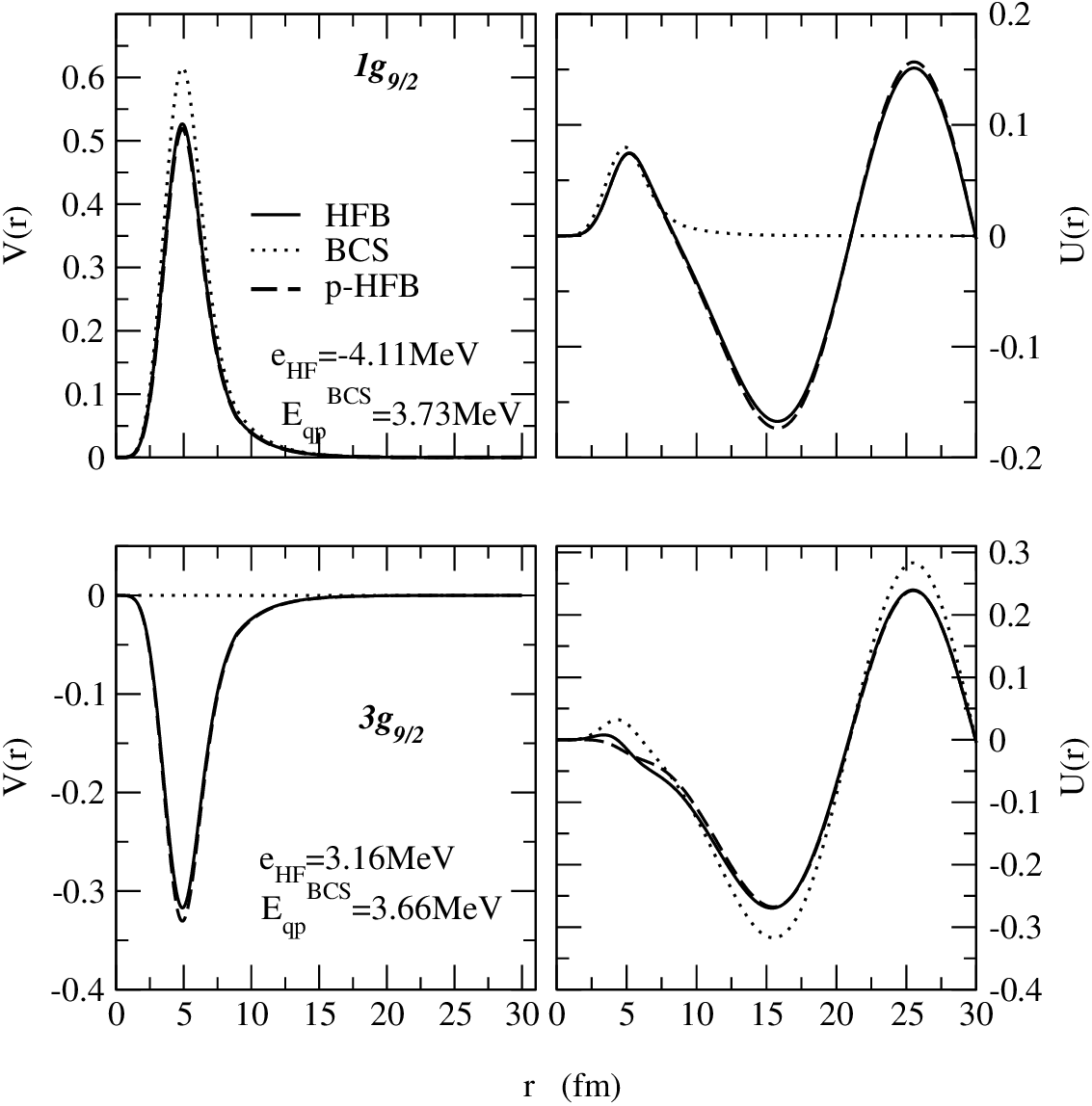}
\caption{Same as Fig.3, but for the 1$g_{9/2}$ (the upper panel) and 
the 3$g_{9/2}$ (the lower panel) states.}
\end{figure}

We next discuss the quasi-particle states which 
have a relatively 
large angular momentum. As an example, Fig. 4 compares the occupation 
probability for the $g_{9/2}$ states obtained by the HFB method (the solid 
line), the BCS approximation (the dotted line) and the perturbative 
approach, p-HFB (the dashed line). 
Again, the p-HFB well reproduces 
the result of the HFB method. Notice that the occupation 
probability for the 3$g_{9/2}$ state at the HF energy of 
$\epsilon=$3.16 MeV is much larger than that in the BCS approximation, which 
essentially leads to a zero occupation probability for this state. 
The unperturbed quasi-particle energy, i.e., the quasi-particle energy 
in the BCS approximation, for this state (3$g_{9/2}$) 
is 3.66 MeV, which is close to 
the quasi-particle energy for the 1$g_{9/2}$ state (3.73 MeV). 
Therefore, according to the perturbation theory, these two states strongly 
couple to each other by the off-diagonal components of the pairing potential. 
The wave functions for these two states are plotted in Fig. 5. Indeed, the 
shape of the wave functions are similar to each other. In the HFB theory, 
hole states appear as narrow resonance states when the quasi-particle 
energy is larger than $-\lambda$ \cite{DFT84,DNW96,B00,BHR03,GSGL01,HM03}. 
The width of the HFB resonance states of this kind originates from the 
coupling between the hole and the particle states due to the off-diagonal 
components of the pairing potential. 
Namely, the particle states which have a similar quasi-particle
energy as the hole state in the BCS 
approximation participate in the resonance state
so that
the quasi-particle wave functions behave similarly between the hole and the 
particle states within the resonance width\cite{U67,GM70,HG04}. 

This important mechanism of the particle-hole couplings is missing 
in the BCS approximation. Note that 
the 3$g_{9/2}$ state is a non-resonant continuum state 
in the absence of the pairing. 
The resonance BCS method, therefore, is not expected to be 
adequate in the circumstance 
where details of wave function plays an important role, for instance in 
the calculation of a strength function with the quasi-particle random 
phase approximation (QRPA) \cite{HS04,Y04}. 

\section{Conclusions}

Within the mean-field approach, the pairing correlation is best incorporated 
with the Hartree-Fock-Bogoliubov (HFB) method. This method consistently 
takes into account the couplings between the particle-hole and the 
particle-particle channels, and thus theoretically robust. 
This method, however, is somewhat cumbersome to solve, 
and the application of the
fully continuum HFB 
calculation has been limited only to spherical systems so far. 
The HFB method, being based on the independent quasi-particle approximation, 
also abandons a simple and intuitive single-particle picture. 
In this paper, we have developed a perturbative model to the HFB 
method, which provides an intuitive connection 
between the HFB and the BCS methods, and at the same time which can be 
solved much easier than the HFB. 
Applying the perturbative approach to weakly bound nuclei, 
we have found that the lowest order 
perturbation reproduces well the results of 
the HFB method both for the quasi-particle energy and the radial 
dependence of wave function. This suggests that the BCS approximation 
provides good unperturbative states, although it leads to inconsistent 
quasi-particle wave functions with the HFB wave functions, 
especially for the smaller component.
As good as the  self-consistent HFB calculation,
one may thus take into account the many-body
pairing correlation by using the perturbative model on top of the
BCS approximation after the convergence of the BCS calculation is achieved.
This scheme might be easily applied to deformed
nuclei for which the self-consitent HFB calculation is very dufficult.

We have also pointed out that non-resonant scattering states in a 
mean-field potential can gain an appreciable occupation probability 
when there is a weakly bound single particle state close to the Fermi 
energy. 
This important coupling effect between particle and hole states 
cannot be incorporated in the BCS 
approximation. Evidently, the BCS approximation is inadequate for the 
QRPA calculations in weakly bound nuclei, 
where the details of the quasi-particle wave functions 
play an essential role. 

\begin{acknowledgments}
We thank N. Sandulescu and Nguyen Van Giai for useful discussions 
on the resonance BCS approach. 
We also 
thank discussions with the members of the Japan-U.S. Cooperative 
Scientific Program ``Mean-Field Approach to Collective Excitations 
in Unstable Medium-Mass and Heavy Nuclei''. 
This work was supported by the Grant-in-Aid for Scientific Research,
Contract No. 16740139 and 16540259, 
from the Japan Society for the Promotions of
Science. 
\end{acknowledgments}

\end{document}